\documentstyle[12pt]{article}
\begin{document}

\title {New approach to deriving gas dynamics equations}

\author{A.V. Kochin\footnote{alexkochin@mtu-net.ru}}

\date {{\it
Central Aerological Observatory\\
141700 Dolgoprudny, Moscow Region\\ Russia}}

%\begin{document}

\begin{titlepage}
\maketitle
\begin{abstract}
We derive the gas dynamics equations considering
changes of velocity distribution function on the scale of a
molecule free path. We define the molecule velocity distribution
function in a specific form so that only molecule velocities after
intermolecular collisions in a chosen fixed volume are taken into
account. The obtained equations differ from the well-known Navier-Stoks
equations and contain the new terms. The continuity equation
includes the second derivatives of dynamical viscosity. The equation of
momentum conservation does not include the volume viscosity and the
equation of energy conservation describes the transformation of gas
mass velocity into the gas thermal energy. One can expect that these new equations
of motion allow to construct a description of macroscopic physical phenomena
in more complicated situations than the standard gas
dynamics equations do.

\end{abstract}
\thispagestyle{empty}
\end{titlepage}

\section {Introduction}

The equations describing gas dynamics are used for studying
various fundamental and applied problems in many areas of
macrophysics. In particular, understanding a number of
concrete macroscopic phenomena is based on solutions to Euler
or Navier-Stokes equations determining the behavior of the gas velocity
field. Standard macroscopic deriving of the equations of gas
dynamics and discussion of their different applications are given,
for example, in \cite{LL}

In contrast with the fundamental Newton dynamic equations, Euler
and Navier-Stokes equations are approximate and provide a
description of a macroscopic gas system in terms of a reduced number of
degrees of freedom. As well known, it is impossible to find an exact
solution to a system of Newton equations of motion for all
interacting gas molecules. Therefore, the equations of gas motion are
derived by methods of statistical mechanics, taking into account
some approximations allowing to develop the description of gas in
terms of macroscopic variables such as gas densities or mass velocities.
One of the most known approaches leading to equations of gas dynamics
in framework of statistical mechanics is the use of
Boltzmann equation. Euler and Navier-Stokes equations are derived from
Boltzmann equation under certain assumptions (viscosity being taken into
account or neglected, respectively).  See for details the ref. \cite{H}.

In its turn, the formulation of Boltzmann equation
is also based on a number of physical hypotheses, which are
discussed in detail, for example, in \cite{H}.
In particular, one of such hypotheses assumes
that for a system, which is close to a locally equilibrium state,
the molecules flying into an arbitrary phase volume have the same
velocity distribution function as the molecules inside this phase
volume. Within the physical meaning, this assumption implies that the molecule
velocity distribution function does not change essentially on a free-path
scale. Therefore, the physical phenomena associated with
the variability of the velocity distribution function on a free-path scale
can be neglected. As a result, deriving of the equations of a gas motion
based on the Boltzmann equation takes into account only those
physical phenomena, where the corresponding velocity distribution function
varies on scale larger than a molecular free path in a gas.

The purpose of this paper is the deriving of the
equations of gas motion for the physical situations, where the
variations of velocity distribution function on a free-path scale
is essential and can not be neglected.
We consider the process of discrete molecule velocity changes,
resulting from intermolecular collisions, as a mechanism responsible for the
variation of the distribution function. The conventional definition of the
velocity distribution function does not take into account molecule collisions
on a free-path scale. Therefore, it will be the same for both a dense gas of
colliding molecules and a highly rarefied gas without collisions. In this
paper, we define a velocity distribution function of molecules in
specific form in such a way that only the molecule velocities
after intermolecular collisions in a chosen fixed volume contribute to
this distribution function.
The velocities of molecules, which pass through the volume without
collisions, are not taken into account. The approach developed in this work,
unlike the one based on Boltzmann equation, does not allow the determination
of an explicit form of the molecule velocity distribution function.
Nevertheless, this method enables deriving the equations of gas motion
under the only assumption that the molecule velocity distribution function
is spherically symmetric.

The paper is organized as follows. Section 2 is devoted to
discussing the standard approach of deriving the gas dynamics
equations. In Section 3 the new definition of molecule
velocity distribution function taking into account only the molecule
velocities after intermolecular collisions is considered. Section 4 is
devoted to the deriving the mass, momentum and energy fluxes for homogeneous gas
on the base of new definition of the velocity distribution function.
In Section 5 we discuss a general method to obtain the
conservation equations for mass, momentum and energy in inhomogeneous
gas. Section 6 is devoted to deriving the concrete form of mass conservation
equation. In Section 7 we derive the momentum conservation
equation. Section 8 is devoted to deriving the energy conservation equation.
In Summary we briefly formulate the basic results of the paper.

\section {Classical principles of deriving the equations of gas dynamics}

Deriving the equations of gas motion by statistical mechanics
methods is built upon several basic principles.

a. The gas fills some domain of space.

b. The gas is in a inhomogeneous state, i.e., density, pressure,
and mass velocity of gas at different points in space are different.

c. It is possible to single out a small enough, but macroscopic domain
of space, which contains a macroscopically large number of
molecules. The changes in density, pressure, and mass velocity of the gas
within the domain are small compared to their changes in the overall
enclosing space.

d. Molecular distribution inside the domain is absolutely
homogeneous. Molecules cross the boundaries of the domain on both sides
thus changing the density, pressure, and mass velocity of the gas .
The molecules flying into the domain increase gas density and pressure
inside it, while the molecules flying out of the domain decrease them.
The mass velocity of the gas in the chosen domain also changes.

e. The rate of change of the parameters depends on the number of
molecules, which fly into or out of the domain, and also on the velocity of
these molecules. It can be calculated from the the conservation laws of
mass, momentum and energy.

f. Mass, momentum and energy of the gas molecules are additive
physical quantities, i.e., if one set of molecules is, on the whole,
characterized by value  $C_1 $ for mass, momentum or
energy and the other by $C_2 $ , then both sets of these
molecules are characterized by value $C_s=C_1+C_2 $.  Such physical
quantities satisfy the local conservation laws for mass, momentum and
energy , which are expressed as follows:
\begin{equation}
\label {1}
\frac {\partial {C}} {\partial {t}} + div\vec {J} _C=0
\end {equation}
Here $C $ is the density of the physical value under consideration (e.g.,
density of mass or density of energy), $ \vec {J} _c $  is
the density of the corresponding flux.  If the physical value is a vector
(e.g., momentum), the following equations must be fulfilled for the
each of its $i$-th component.
\begin {equation}
\label {2} \frac {\partial {C} _i} {\partial {t}} +
\frac {\partial} {\partial {x} _j} \hat {J} _ {i j} =0
\end {equation}
Thus, the density of a flux for a scalar quantity (mass, energy) will be a
vector, and the one for a vector quantity (momentum) will be a second-rank
tensor.

Equations ~ (\ref {1}) and~ (\ref {2}) are valid for an arbitrary domain
and if no sources of mass, energy or momentum are inside the
domain, these equations allow us to construct a description
of physical effects on the base of those assumptions, which are
incorporated in the conservation law equations for density of
mass, momentum and energy fluxes.

In their turn, the density of mass, momentum and energy fluxes in a gas
can be calculated from the velocity distribution function $f (\vec
V, \vec r, t) $. The equations for the density of mass, momentum and
energy fluxes have the following form:

the i-th component of the density of flux $J _ {C} ^ {i} $
for the physical value $ C $ (mass, momentum, energy)
is equal to
\begin {equation}
\label {3}
J _ {C} ^ {i} = \rho\int _ {-\infty} ^ {\infty} dV_x\int _ {-\infty} ^ {\infty}
dV_y\int _ {-\infty} ^ {\infty} {dV_z} f (\vec {V}, \vec {r}, t) C {V_i} =
\rho\int f (\vec {V}, \vec {r}, t) C {V_i} d \vec {V}
\end {equation}
where $ \rho $ is the density of gas at point $ \vec {r} $
at time $t $, $V_i $ is the i-th component of the total
velocity of molecules, $f (\vec {V}, \vec {r}, t) $ is the normalized
function of molecule velocity distribution
\begin {equation}
\label {3} \int _ {-\infty} ^ {\infty} dV_x\int _ {-\infty} ^ {\infty}
dV_y\int _ {-\infty} ^ {\infty} {dV_z} f (\vec {V}, \vec {r}, t) =1
\end {equation}

Hence, if the molecular velocity distribution function is given, the
calculation of the physical quantities is reduced to formal
procedures.  However, it seems that it is impossible to find the function
$f (\vec V, \vec r, t) $
only from experimental data. Therefore to define this function, we
have to use a theoretical approach. The generally accepted definition
of the distribution function is formulated as follows.

{\bf DEFINITION 1 (SNAPSHOT)}

The function $f (\vec V, \vec r, t) $ is given by
a velocity histogram of the molecules in volume dV. Hereafter, the
method, which is generally used to find this function will be referred
to as "instant
snapshot". The virtual camera located inside volume dV
centered around point $ \vec r $ records at time t all the molecules inside
this volume. It is assumed that this camera also measures and records the
velocities of all the molecules inside this volume at a given time t.
Then one counts the number of molecules with a certain velocity vector
(velocity modulus and direction) and then calculates the velocity histogram
of the molecules.

After that the following formal transformations are fulfilled. The mass
velocity $ \vec {u} $ for all molecules in the domain under
consideration is determined in the form

\begin {equation}
\label {4}
\vec {u} = \int {f (\vec {V}, \vec r, t) \vec {V} d\vec {V}}
\end {equation}
A difference between the total velocity of a molecule $ \vec V $ and the
mass velocity $ \vec u $ is called proper velocity of a molecule in
the moving with gas reference frame and denoted $ \vec v $. The proper
velocity is called usually a thermal velocity.  For proper
velocity, the velocity distribution function $ \tilde {f} (\vec v, \vec
r, t) $ is also introduced, it is evident that this function is
functionally connected to distribution function
velocities $f (\vec V, \vec r, t) $ on total velocities.

Now, for a final derivation of the macroscopic equations of motion
we have to define a form of distribution functions of
molecules for the proper velocities. For this purpose the
following assumption is usually introduced: the gas in considered domain
has a locally equilibrium state. In this case the
proper velocity distribution function is well-known Maxwell
distribution function.  After that, a formal consideration based on
eqs.~ (\ref {1}),~ (\ref {2}) and~ (\ref {3}) leads to the
of Euler equation ( see the details in \cite{H}).

To derive the Navier-Stokes equation, we have to represent the
distribution function as the sum of the locally
equilibrium function and some small unknown function (small
perturbation). Then, taking into account the certain assumption about
the small perturbation we can find the corrections to the Euler
equation describing the effects
associated with viscosity and thermal conductivity (Navier-Stokes
equation; see the details in \cite{H}).

\section {Definition of proper velocity
distribution function taking into account the molecular collisions}

The {\bf DEFINITION 1} does not take into account the
collisions of molecules and will be the same both for gas without
collisions, and for gas with collisions. If our aim is to study
the physical values in a gas for the situations where molecular
collisions are essential, we have to find a distribution
function containing the effects of molecular collisions in its
expression from the very beginning. Therefore, we change the above
{\bf DEFINITION 1} as follows.

Let's consider a domain filled with gas of
colliding molecules and let the free path
$\lambda$ be much smaller then the characteristic scale of the domain and much
smaller then the characteristic scale of the problems under consideration.
These assumptions are fulfilled for the most practically
important problems. For example, free path  of molecules in air
at atmospheric pressure and room temperature is about $ 10 ^ {-7} $ cm
whereas the characteristic scale of any macroscopic problem is
much larger.

To find the distribution function taking into account the molecular
collisions we accept the following definition.

{\bf DEFINITION 2 (CINEMA FILM)}

As before, we single out a domain $dV $ which size  is
small comparing to the free path of molecules in a
gas. When we have considered the velocity histogram in {\bf DEFINITION
1}, the velocities of molecules have been shot by the camera.  Now
we use a movie camera. It shoots on a film a series of consecutive
pictures of molecules in the domain $dV $ during the time interval dt,
duration of which will be defined later.

Let us divide the molecules, whose images are shot on the picture
area of a film into two groups. The first group contains the molecules,
which have taken part in the collisions inside the considered domain.
Now we assume that the velocities for distribution function are
taken into account only for molecules right after the collision.
As a graphic illustration for such a
situation we propose that colliding molecules in the volume
$dV $ under consideration are visible and can be seen by the observer,
while the molecules flying without collisions are invisible.  The observer
sees only painted molecules
and, therefore, only their velocities define the
distribution function for him. Second group of molecules,
which pass through the above volume $dV $ without collisions, does
not contribute to distribution function. We will define the
normalized molecular velocity distribution function $ f_s (\vec V, \vec
r, t) $ on the base of set of velocities obtained in such a
way. Namely this function will be further considered as the molecular
velocity distribution function $ f_s (\vec V, \vec r, t) $ at the given
point of space, at the given time t.

In order to define this function statistically, in the same way
as the function $f (\vec V, \vec r, t) $ according to the
{\bf DEFINITION 1} it is necessary to take into account that a number
of molecules, which is used to evaluate of the molecule velocity
distribution function, would be equal to a total number of molecules in
volume dV at the time t. The number of collisions is
proportional to time.  If the time interval dt is approximately $
\lambda/\bar {v} $, where $ \bar {v} $ is a average modulus of proper
velocity, a number of colliding molecules will be equal to a number of
molecules in chosen volume $ dV $.  For gas under normal conditions
this time is about $ 10 ^ {-11} s$ and is much less then characteristic
time of change of macroscopic parameters of gas.  Since the
distribution function is macroscopic characteristic of gas, the
times t and t+dt are not distinguished one from another. Therefore,
we can treat the function determined according to the above rules as
molecule velocity distribution function at the time t. We
will call this statement the {\bf DEFINITION 2}.

One has to note that the above definition does not allow to obtain an explicit
form of the distribution function in the same sense as in the
{\bf DEFINITION 1}. Nevertheless, it is possible a priori to draw a
conclusion about the important features of such a function.

Let us introduce the definition of the mass velocity $
\vec u $ and proper velocity $ \vec v =\vec V - \vec u $ as above.
According to the {\bf DEFINITION 2}, the proper velocity of a molecule
is  the velocity of a molecule after collision in
a reference frame moving with gas. The vectors of proper velocities of
molecules after collisions have equiprobable directions in this reference
frame. Therefore, the proper velocity
distribution function $ \tilde {f} _s
(\vec v, \vec r, t) $ will exhibit a spherical symmetry. This property of
spherical symmetry means that all directions of proper
velocity vector of molecules $ \vec v $ are equivalent.

The property of spherical symmetry is very important to find the
explicit expression for macroscopic quantities according to the
equation (3). Taking into account the spherical symmetry one define
the dimensional distribution function function depending
only on modulus of proper velocity $ | \vec v | $,
 $ \tilde {f}_s = \tilde {f} _s (| \vec v |) $. Then, the
fluxes of mass, momentum and energy can be easily calculated.

To simplify the expressions we will write the
distribution function of the modules of proper velocity
as $f_s (v) $. The normalization condition for this function $f_s (v)$
looks like
 \begin {equation}
\label {5} \int _ {0} ^
{\infty} {\tilde {f} _s (| \vec {v} |) d |\vec {v} |} = \int _ {0} ^
{\infty} {f_s (v) dv} =1
\end {equation}
Namely the function $\tilde {f}_s$ will be used further to derive
the macroscopic equations of motion.

\section {Equations for flux of mass and momentum in homogeneous
rest gas.}

Volume  filled with gas can be treated as continuous up to spatial
scale of free path of molecules. It means,
that only the collisions cause a change of velocity vector of a molecule.
After collision, the velocity of the molecule does not vary,
until it collides with another molecule.
Between the collisions, the molecule passes an average distance
approximately equal to free pass $ \lambda $. The molecule, which
had a collision, keeps a constant velocity during the free
path (of course, if there is no external field).

Hence, the gas of colliding molecules can be considered
as the set of points, each of them is spherical
"source" of molecules with velocity distribution function belonging
to the given concrete point. Intensity of a molecular flux from
each such a point decreases proportionally to an inverse square
distance from the given point. Let us consider a sphere of the radius
$ \lambda $ surrounding the radiating point. It is clear that
beyond this sphere,
the intensity of a "source" becomes equal to zero,
since there should be a next collisions of molecules here. After
the collisions, the velocities of molecules belong to other distributions,
which appropriate to other "radiating" points located
on a surface of the same sphere.

The intensity of a considered point "source" is
a product of velocity of molecules and their density. The above
definition of velocity distribution function
does not allow to find a
density of colliding molecules in the volume under consideration. This
density (we denote  it $ \rho ^ {\prime} $) can be
obtained using
the following arguments. Let us introduce a surface into a gas and
an observer behind this surface. This observer counts a number of
molecules flying through the surface with different velocities.
It is found that N molecules have a flux under the certain
angle $ \theta, \varphi $ with the certain velocity V.
It is not possible to say, from which of radiating points one or other
molecule has started its flight.
The observer is able to determine only a sum of intensities from all sources
 and this sum
is equal to a product of the true density of the gas $ \rho $
near the surface and
the velocity of molecules. On the other hand, this sum is equal to the integral
of the intensity of all radiating points at distances from 0 up to $ \lambda $
from a point of observation. Therefore, one can write
\begin {equation}
\label {6}
\rho V (\theta, \varphi) = \int _ {0} ^ {\lambda} \rho ^ {\prime} V (\theta,
\varphi) dl
\end {equation}
Hence
\begin {equation}
\label {7}
\rho ^ {\prime} = \frac {1} {\lambda} \rho
\end {equation}

Let's place in homogeneous
non-moving gas of colliding molecules a surface $dS$ located at a point
 with the coordinates $x,y,z$,
and small comparing to $\lambda^2$.
Normal vector to this surface
is directed along the axis X. Only the molecules, which had collisions inside
the sphere of radius $\lambda$, cross such a surface.
The molecules which had the collisions in hemisphere located to the left from $dS$
fly along the positive direction of X axis and the molecules which had the
collisions in hemisphere located to the right from $d$S fly along the negative direction of
X axes. The total flux through
dS will be equal to the difference of two above fluxes.

The flux is considered at some time t. At this time point,
only the molecules which had the collisions near the surface $dS$
and the molecules which had the collisions on the distance $l$ from
the surface $dS$ fly through
the given surface. Therefore, the distribution functions
belong to the different time points. However, maximal difference in time
is equal to $ \lambda/V $, that for gas at normal
conditions is around $ 10 ^ {-11} $ s. Hence, one can neglect
a difference in time between collisions
and to consider, that all molecules had collisions
at the same t (simultaneously).

Let's calculate a flux along positive
directions of the axis Õ. We set a polar coordinate system
in the hemisphere,
where a distance from $dS$ to a point inside a hemisphere is equal to $r $,
and the angle between normal to $dS$ and direction of $r$ is equal
to
$ \theta $.
Let's divide the hemisphere
into layers parallel to $dS$ and located at the various distances $l$ from
$dS$.
A thickness of a layer is $dl$. A distance from $dS$ to a point
at layer at the
distance $l $ is equal to $r=l \,/cos\theta $.  We associate a small volume
$dV$ with each point of hemisphere attached to $dS$ and consider this
volume $dV$ as a source of molecules with
distribution on modules of proper velocity $f_s (v) $. The intensity of
source for
molecules with proper velocity from v to v+dv
is equal to $ \rho f_s (v) vdvdV/\lambda $.
Density of a flux of molecules $dI$
decreases proportionally to inverse square of distance
$ r^2=l^2/cos^2\theta $. The flux $dJ _ {\rightarrow} $ through $dS$ is
equal to
the product density $dI$ and a projection $dS$, which is equal to
$ dS\cos\theta $. Thus, flux $dJ _ {\rightarrow} $ from a single
radiating point can be written as follows
\begin {equation}
\label {6}
dJ _ {\rightarrow} = \, {\frac {\rho \, f_s (v) vdvdV {\it dS} \, \cos^3 {\it
\theta}} {4
\pi \, {l} ^ {2} \lambda}}
\end {equation}
Let's divide a hemisphere of radius $ \lambda $ into layers
of thickness $dl$,
located at the distance $l $ from a surface $dS$.
Each layer, in its turn, is divided into the rings
with constant angle $ \theta $,
where radius of the ring $R=l \, tan\theta $. Volume of the ring is $RdRdl $.
Also, $dR=l \, d\theta/cos^2\theta $.
The flux $dJ _ {c\rightarrow} $ through a surface $dS$ from such a ring
is equal to
\begin {equation}
\label {7}
dJ _ {c\rightarrow} = \frac {\rho f_s (v) vdv\cos\theta
dS\tan\theta {d\theta} dl} {2\lambda}
\end {equation}
For all rings belonging to a layer at the distance $l $, the angle $ \theta $
varies from 0 to $arccos (l/\lambda) $. Therefore, the flux from the whole layer
$ dJ _ {l\rightarrow} $ is equal to integral from $dJ _ {c\rightarrow} $
over
$ \theta $ from 0 to $arccos (l/\lambda) $. As a result we get
\begin {equation}
\label {8}
dJ _ {l\rightarrow} = \int _ {0} ^ {arccos (l/\lambda)} dJ _ {c\rightarrow} \, \
d\theta = \frac {\rho \,\ f_s (v) vdv dS\left (\lambda-l\right) dl} {
2 {\lambda} ^2}
\end {equation}
The flux through $dS$ is equal to the sum of fluxes from all layers,
which are located away
from $dS$ at the distance from 0 to $ \lambda $.
\begin {equation}
\label {9}
J _ {m\rightarrow} = \int _ {0} ^ {\lambda} dJ_l
dl =\frac {1} {4} \, \rho \, f_s (v) vdv {\it dS}
\end {equation}
If we divide the expression obtained for any surface $dS$ by
$dS$ we will have the expression for a mass flux density
along axis Õ for the
molecules, whose velocity is equal to v. The integration over v gives
mass flux density along a positive direction
of X-axis as a result of the proper velocity
\begin {equation}
\label {10}
J _ {mx\rightarrow} = \int _ {0} ^ {\infty} dv J_m
= \frac {1} {4} \, \rho \,\bar {v}
\end {equation}
Together with the flux of mass along a positive direction of an X-axis
there exists a flux of mass coming along a
negative direction of
axes Õ and having the same value. The resulting flux of mass is equal
to a difference between two above fluxes, i. e. the total flux is equal to
zero.

X-component of the momentum of the each individual molecule is equal to
$ mvcos \,\theta $. Multiplying ~ (\ref {6}) with $vcos \,\theta $
we obtain an
equation for density of a flux for X-component of momentum from
each radiating point. The similar calculations
lead to the following result for the density of the flux X-component of
the momentum
along a positive direction of X-axis
\begin {equation}
\label {11}
J _ {px\rightarrow} = \frac {1} {6} \, \rho \,\bar {v^2}
\end {equation}

Apart, a flux coming along a negative direction of an X-axis
also exists. It has the same absolute value but opposite direction. The
resulting density of a flux of X-component
of the momentum is equal to a difference of above two fluxes. However the
momentum, in
contrast with a mass, is a vector quantity. Therefore, a  difference of the
fluxes of
momentum along positive and along negative directions
of X-axis is equal to the doubled value of ~ (\ref {11}).
As a result, we obtain the known formula for hydrostatic pressure
\begin {equation}
\label {12}
P = \frac {1} {3} \, \rho \,\bar {v^2}
\end {equation}
For density of a flux of energy along a positive direction of an X-axis
we have
\begin {equation} \label {13}
J_Ex =\frac {1} {8} \, \rho \,\bar {v^3}
\end {equation}
The energy is scalar quantity, therefore the resulting
density of a flux is equal to zero, like for density of a flux of mass.

   \section {General approach of deriving the equations of motion for
   inhomogeneous gas.}

Let's consider, as in the previous section, a domain  filled with
gas, where a
free path is much smaller than the size of the domain
and much smaller than the characteristic scale of a problem.
The gas is in a inhomogeneous state, i. å., a density and
velocity distribution function at different points of
the volume differ from each other.

Let's place any surface in gas. This
surface will be crossed by molecules which come from the both
sides of the surface.
Since the molecules do not change the velocity on the scale $ \lambda $,
the effective sources of fluxes of molecules is located
from each other at the distance determined by free path
$ \lambda $. It means, the fluxes of molecules on the different
sides
of the surface  will have the different parameters: density,
temperature, mass velocity and the velocity distribution function.
The difference of physical parameters of molecular fluxes leads to
appearance of terms with spatial derivatives
in expressions for fluxes for mass, momentum and energy.

The derivation of the equations is carried out under the following
restrictions:

1. The linear approximation is considered. Accordingly to this, only the
first spatial derivatives of density, mass and proper velocities  are
taken into account.

2. The approximation of small velocities is considered. It means,
 a module of the mass velocity $ | \vec u | $ is
 much
 smaller than average module of proper velocity $ \bar {|v |} $.

  \section {The mass conservation equation.}

To derive the mass conservation equation, taking into account the collisions
of molecules,
it is necessary to find an expression for mass flux, suitable for these
conditions.
Let's associate with a point $x,y,z$ a small surface $dS$, directed
perpendicularly
to X-axis in non-moving laboratory reference frame.

At the given point $x, y, z $ gas is characterized by
the density $ \rho $, mass velocity $ \vec u $ and distribution function on
modules of proper velocity $f_s (v) $.

The calculations for inhomogeneous gas are basically similar to
the calculations
for homogeneous gas. The sphere attached to $dS$
is divided into layers and etc.
Equation (6)is written as follows
\begin {equation}
\label {14}
dJ _ {\rightarrow} = \, {\frac {\rho \, f_s (v) (v \cos {\it \theta} +u_x)
dv {\it dS} \,
\cos^2 {\it \theta}} {4
\pi \, {l} ^ {2}}}
\end {equation}
where $u_x $ is X-th component of the mass velocity.
The flux from a ring is written in
appropriate way
\begin {equation}
\label {15}
dJ _ {c\rightarrow} = \frac {\rho
f_s (v) dv (v\cos\theta+u_x) dS\tan\theta {d\theta}} {2}
\end {equation}
For all rings belonging to the layer at the distance $l $, angle $ \theta $
varies from 0 to $arccos (l/\lambda) $. Therefore, the total flux from
all layer
$ dJ _ {l\rightarrow} $ is equal to integral from $dJ _ {c\rightarrow} $
over
$ \theta $ from 0 to $arccos (l/\lambda) $.
\begin {equation}
\label {16}
dJ _ {l\rightarrow} = \int _ {0} ^ {arccos (l/\lambda)} dJ _ {c\rightarrow} \, \
 d\theta = \frac {\rho \,\ f_s (v) dv dS\left (v (\lambda-l) -u_x \ln\left (\frac {l} {\lambda}
 \right) \right)} {2\lambda}
\end {equation}
The flux through $dS$ is equal to the sum of the fluxes from all layers,
which are located away
from $dS$ at the distances from 0 to $ \lambda $. Since the infomogeneous gas
is considered, the density $ \rho $, proper
velocity $v $ and X-th component of the mass velocity $ u_x $
are the function of the distance from $dS$ to the layer at the distance $l$.
In first approximation we restrict ourselves
to the linear terms in spatial derivatives in expansion of all
quantities.
\begin {equation}
\label {17}
\rho (l) = \rho-\frac {\partial \rho} {\partial {x}} l
\end {equation}
\begin {equation} \label {18}
v (l) =v-\frac {\partial v} {\partial {x}} l
\end {equation}
\begin {equation} \label {19}
u_x (l) =u_x-\frac {\partial u_x} {\partial {x}} l
\end {equation}
Partial derivative on y and z do not contribute to
the expressions for
flux on axis Õ, since they are averaged under integration over
ring.

The signs at partial derivatives in eqs. (17), (18) and (19)
are defined by a direction of a flux. It means, the sign for a flux
on left side from $dS$ will be opposite to sign on right.

The flux of mass through $dS$ is equal to the difference of fluxes on the
left and on the right sides. Since
the case $ | \vec u \ | <<\bar |v | $ is considered,
the terms containing the powers of  $ \lambda $  higher than first
and the products of the quantities
$ u_x\lambda $, which are small comparing to $v\lambda $, are omitted.
Expression for
X-th component of a resulting vector of a mass flux resulted from
the
molecules with velocity v can be written in the following form
\begin {equation}
\label {20}
  J _ {mxv} = \rho f_s (v) dv\left (\rho u_x-\frac {1} {6} \rho\lambda\frac {\partial v} {\partial {x}}
  -\frac {1} {6} v\lambda\frac {\partial \rho} {\partial {x}} \right)
\end {equation}
For obtaining the final result, it is necessity to integrate eq.(20)
over velocity.  It is reasonable to assume, that the form
of distribution function  $f_s (v) $ does not change on scale $ \lambda $,
i. e. relative changes of individual velocities
are proportional to changes of average velocity. Then we can
write the relation
\begin {equation}
\label {21}
  \frac {\partial v} {\partial {x}} = \frac {v} {\bar {v}} \frac {\partial \bar
  v} {\partial {x}}
\end {equation}
The equation (20) after integration over velocities
looks like

\begin {equation} \label {22}
  J _ {mx} = \rho f_s (v) dv (\rho u_x-\frac {1} {6} \lambda\frac {\partial \left (\rho \bar v
  \right)} {\partial {x}}
\end {equation}
By definition, a product of the density, free path and
the average velocity is proportional to the viscosity $ \mu $
 \begin {equation}
 \label {23}
\mu =\frac {1} {3} \rho\bar v\lambda
\end {equation}
Then the relation (22) can be written as follows
\begin {equation} \label {24}
  J _ {mx} = \rho u_x-\frac {1} {2} \frac {\partial \mu} {\partial {x}}
\end {equation}

The expressions for Y-th and Z-th components
of a vector of a mass flux density are written in the similar way
\begin {equation}
\label {25}
  J _ {my} = \rho u_y-\frac {1} {2} \frac {\partial \mu} {\partial {y}}
\end {equation}
\begin {equation} \label {26}
  J _ {mz} = \rho u_z-\frac {1} {2} \frac {\partial \mu} {\partial {z}}
\end {equation}
Expression for the vector of mass flux density looks like
 \begin {equation}
 \label {27}
J_m =\rho \vec u-\frac {1} {2} grad\mu
\end {equation}
Using eqs. (1) we get the mass conservation equation in
the form
\begin {equation}
\label {28}
\frac {\partial\rho} {\partial {t}} +div (\rho\vec
u) -\frac {1} {2} div (grad\mu) =0
\end {equation}
The expression (28) differs from the known continuity equation
in the derivatives of viscosity. Factor 1/2 at the corresponding
term is explained by hemispheres attached to each other which are the
integral sources of molecules.

\section {The momentum conservation equation}

The momentum conservation equation is derived in the similar way to mass
conservation equation with the help of substituting the appropriate
quantities for momentum components
by the untegrand of eq.(2). Density of a flux of X-th component
of momentum along the X-axis $dj _ {px\rightarrow} $ from the individual
radiating
point is equal to

\begin {equation}
\label {29}
dj _ {px\rightarrow} = \, {\frac {\rho \, f_s (v) (v \cos {\it
\theta} +u_x) ^2 dv {\it dS} \,
\cos^2 {\it \theta}} {4
\pi \, {l} ^ {2}}}
\end {equation}
Density of the flux of X-th components of a momentum from a layer
at the distance $l$
$ dJ _ {lpx\rightarrow} $ has the form
\begin {equation}
\label {30}
dJ _ {lpx\rightarrow} = \frac {\rho \,\ f_s (v) dv dS
 \left [{v} ^ {2} {l} ^ {2} +4 \, vul\lambda+2 \, {u} ^ {2} \ln ({\frac {l} {\lambda}}) {\lambda} ^ {2} - {
v} ^ {2} {\lambda} ^ {2} -4 \, vu {\lambda} ^ {2} \right]} {{\lambda} ^ {2}}
\end {equation}
Flux from whole left hemisphere is equal to integral from
$ dJ _ {lpx\rightarrow} $ over $l$ from 0  to $ \lambda $. The flux
from the right hemisphere $dJ _ {lpx\leftarrow} $ will be calculated
by the same manner.
The resulting flux through $dS$ is equal to the sum of these fluxes and
for fixed velocity v can be written as follows
\begin {equation}
\label {31}
J _ {Pxxv} =f_s (v) dv\left [\frac {1} {3} \rho v ^ {2} + \rho
{u_x} ^ {2} -\frac {1} {3}
\lambda \left (\rho u_x\frac {\partial v} {\partial x} +
v u_x\frac {\partial \rho} {\partial x} +
 \rho v \frac {\partial u_x} {\partial x} \right) \right]
\end {equation}
After integration over v we get the expression for density of a flux
of
X-th component of a momentum along the X-axis
\begin {equation}
\label {32}
J _ {Pxx} = \frac {1} {3} \rho \bar {v ^ {2}} + \rho {u_x} ^ {2} -\frac {\partial
\mu u_x} {\partial x}
\end {equation}
Density of a flux of Y-th component of a momentum along the X-axis
$dj _ {py\rightarrow} $ from a single radiating point is equal to

\begin {equation}
\label {33}
dj _ {px\rightarrow} = \, {\frac {\rho \, f_s (v) (v \cos {\it
\theta} +u_x) u_y dv {\it dS} \,
\cos^2 {\it \theta}} {4
\pi \, {l} ^ {2}}}
\end {equation}
The calculations for density of a flux of Y-th component along the X-axis
lead to the following expression
\begin {equation}
\label {34}
J _ {Pxy} = \rho {u_x} {u_y} -\frac {1} {2} \frac {\partial \mu
u_y} {\partial x}
\end {equation}
Expression for density of a flux of Z-th component of a momentum along the
X-axis is equal to

\begin {equation}
\label {35}
J _ {Pxz} = \rho {u_x} {u_z} -\frac {1} {2} \frac {\partial \mu
u_z} {\partial x}
\end {equation}
Other components of a vector of a flux of a momentum are calculated
in the same way. After substituting the expressions obtained into
eq. (2) and subtracting from the result the conservation equation for mass
(24),
multiplied with $ \vec u $, we get the following form of the
momentum conservation equation
\begin {equation}
\label {36}
  \rho \,\left ({\frac {\partial} {{\it \partial t}}} +u {\it \nabla} \right) u + {\it
  grad} \Pi
- {\it grad} (\mu) {\it nabla} (u) - {\it D} =0
\end {equation}
where the vector of pressure $ \Pi $ is sum of hydrostatic pressure $Ð$ and
a vector of hydrodynamical pressure. This pressure $\Pi$ is expressed in the
following form

\begin {equation}
\label {37}
  \Pi =\begin {array} {c}
  P-\mu\it\frac {\partial u_x} {\partial x} \\
  P-\mu\it\frac {\partial u_y} {\partial y} \\
  P-\mu\it\frac {\partial u_z} {\partial z}
  \end {array}
\end {equation}
The parameter D in eq (39) is written as

\begin {equation}
\label {38}
 D = 1/2 \,\mu \left [\begin {array} {c}
 {\frac {\partial ^ {2} \it u_x} {
\partial {x} ^ {2}}} + {\frac {\partial ^ {2} \it u_x} {\partial {y} ^ {2}}}
+ {\frac {\partial ^ {2} \it u_x} {\partial {z} ^ {2}}} \\ {\frac {\partial
^ {2} \it u_y} {\partial {x} ^ {2}}} + {\frac {\partial ^ {2} \it
u_y} {\partial {y} ^ {2}}} + {\frac {\partial ^ {2} \it u_y} {\partial
{z} ^ {2}}} \\ {\frac {\partial ^ {2} \it u_z} {\partial
{x} ^ {2}}} + {\frac {\partial ^ {2} \it u_z} {\partial {y} ^ {2}}} + {\frac
{ \partial ^ {2} \it u_z} {\partial {z} ^ {2}}}
\end {array} \right]
\end {equation}
Appearance of the hydrodynamical pressure in the expression for
pressure $\Pi$ is explained by the following way.
The pressure is defined as a
flux of a momentum of molecules colliding with some surface.
Quantitatively, this flux of momentum
is equal to a product of momentum $mV$ of individual molecule with
flux of number of molecules $nV$ colliding with the surface. Then, we get
for hydrostatic pressure $P\sim <mVnV> = \rho <V^2> $.
The mass velocity leads to the change of both values of the momentum
and the number of molecules, colliding with the surface. However,
this effect works only at the distance of free path.
Therefore, the correction to the hydrostatic pressure is equal to a product mV
with
$nV\lambda {\frac {\partial u_i} {\partial {x_i}}} $, that is equal
to
$ \mu {\frac {\partial u_i} {\partial {x_i}}} $. Thus,
expression (37) defines the actual measurable pressure on the
surface, oriented along axes X, Y and Z, respectively.

\section {The energy conservation equation.}

Density of an energy flux from the radiating point is equal to
\begin {equation}
\label {39}
J _ {Ex} = \frac {\rho \, f_s (v) (v \cos {\it
\theta} +u_x) (v^2+2v \, u_x \, cos\theta + {\vec u} ^2) dv {\it dS} \,
\cos^2 {\it \theta}} {8
\pi \, {l} ^ {2}}
\end {equation}
After calculation, we get the following expression for the resulting
density of the energy flux along the X-axis
\begin {equation}
\label {40}
J _ {Ex} = \frac {1} {2} \ {, \it ux} \rho\left (\frac {5\bar {v^2}} {3} + {u} ^ {2}
\right) -\frac {1} {12} \frac {d (\rho\lambda\bar {v^3})} {dx} -\frac {3} {4}
\frac {d (\mu {u_x} ^ {2})} {dx} -\frac {1} {4} \frac {d (\mu {u_y
} ^ {2})} {dx} -\frac {1} {4} \frac {d (\mu {u_z} ^ {2})} {dx}
\end {equation}
Second term in the right hand side of the obtained expression represents
the partial derivative of $ \frac {1} {12} \rho\lambda\bar {v^3} $.
Let's transform it as follows. For an any
velocity distribution function one can write
\begin {equation}
\label {41}
  \frac {1} {12} \rho\lambda\bar {V^3} = \frac {1} {12} \rho\lambda\bar {V} \bar {V^2} k (\theta)
  = \frac {1} {12} k (\theta) \mu\bar {V^2} =K\bar {V^2}
\end {equation}
Here the parameter $K $ can be interpreted as gas thermal conductivity
(we point out that the $K$ depends on the temperature as well as the
coefficient of viscosity $ \mu $.). In
accordance to that,
the expression for the vector of the energy flux takes the form
\begin {equation}
\label {42}
  J_E =\begin {array} {c}
  \frac {1} {2} \ {, \it u_x} \rho\left (\frac {5\bar {v^2}} {3} + {u} ^ {2}
\right) -\frac {d (K\bar {V^2})} {dx} -\frac {3} {4}
\frac {d (\mu {u_x} ^ {2})} {dx} -\frac {1} {4} \frac {d (\mu {u_y
} ^ {2})} {dx} -\frac {1} {4} \frac {d (\mu {u_z} ^ {2})} {dx} \\
\frac {1} {2} \ {, \it u_y} \rho\left (\frac {5\bar {v^2}} {3} + {u} ^ {2}
\right) -\frac {d (K\bar {V^2})} {dy} -\frac {3} {4}
\frac {d (\mu {u_y} ^ {2})} {dy} -\frac {1} {4} \frac {d (\mu {u_x
} ^ {2})} {dy} -\frac {1} {4} \frac {d (\mu {u_z} ^ {2})} {dy} \\
\frac {1} {2} \ {, \it u_z} \rho\left (\frac {5\bar {v^2}} {3} + {u} ^ {2}
\right) -\frac {d (K\bar {V^2})} {dz} -\frac {3} {4}
\frac {d (\mu {u_z} ^ {2})} {dz} -\frac {1} {4} \frac {d (\mu {u_x
} ^ {2})} {dz} -\frac {1} {4} \frac {d (\mu {u_y} ^ {2})} {dz}
\end {array}
\end {equation}
The conservation equation for energy is obtained after
substituting eq. (42) into the eq. (1) and the
subsequent subtraction from the result of
the mass conversation equation (28), multiplied with $u^2/2 $,
 and momentum conservation equation (36), multiplied with a vector
of mass velocity $ \vec u $.
 \begin {equation}
 \label {43}
\left (\frac {\partial} {\partial {t}} + \vec {u} \nabla\right) P +\frac {5} {3} Pdiv\vec {u} -
\nabla^2 (K\bar {V^2})
-\mu/3\sum_i\sum_j\left (\frac {\partial \it u_i} {\partial {x_j}} \right) ^2-
B=0
\end {equation}
Where $B$ is equal
\begin {equation} \label {44}
  B=2\mu/3\sum_i\left (\frac {\partial \it u_i} {\partial
  {x_i}} \right) ^2+2/3\sum_i \left (u_i\frac {\partial \it u_i} {\partial {x_i}}
  \frac {\partial \it \mu} {\partial {x_i}} \right) + \mu/3\sum_i \left (u_i\frac
{ \partial^2 \it u_i} {\partial {x_i} ^2} \right)
\end {equation}
First three terms in the equation (46) are similar
to the known energy conservation equation taking into account
the process of heat transfer. Two last terms in the equation (46)
describe a process of energy dissipation of the ordering motion
into the thermal energy of gas due to the work of forces of viscosity.
Energy dissipation can be understood as follows.
There are the collisions of molecules at each space point.
The molecules flying to the given point
have the different velocity distribution, in particular they have the
different mass velocities. After the collision, the molecules belong to
the same velocity distribution.
As a result, a part of energy of the ordering motion approximately
equal to
$ \mu\left ({\frac {\partial u_i} {\partial {x_j}}} \right) ^2 $
will be always
used for heating of gas. Hence, all processes in a gas,
resulted from the change of the mass velocity, are
irreversible processes.
Thus, the equation (46) leads to the possibility
to describe correctly the energy dissipation at the gas motion.
\\

\section {Summary}

We have introduced the new definition of the molecule velocity
distribution function describing the molecular motion on a
free-path scale. Using this definition we derived the equations
of gas dynamics containing the new terms in comparison with well
known equations ref. \cite{H}. The equation of mass conservation (31) includes the new
contribution with second derivative of dynamic viscosity.
The equations of momentum conservation (39 - 41) do not contain the volume
viscosity. The equations of energy conservation (46 -47) describe the
transformation of energy of gas mass velocity into the thermal energy of gas.

The new equations of gas dynamics derived here are approximate like
the Euler and Navier-Stokes equations.
We have restricted ourselves by the approximation of small velocities.
Nevertheless, we hope that these new equations will be useful for
description of the nonequilibrium phenomena in various areas of
macroscopic physics. For example, these equations can be
applicable to computation of atmospheric processes like turbulence and
solution of the aerodynamic problems like modeling a subsonic fluxes.
 We plan to discuss such problems on the base of our new
equations in forthcoming works.

{\bf Acknowledgements.}
I am very grateful to I.L. Buchbinder for help in work and
useful discussions. Also, I would like to gratitude V.L. Kuznetsov
for discussions on various aspects of gas dynamics.


\begin{thebibliography}{000}

\bibitem{LL} L.D. Landau, E.M. Lifshits, Hydrodynamics, Nauka,
1986 (in Russian)

\bibitem{H} K. Huang, Statistical Mechanics, John Wiley and Sons,
1963

\end{thebibliography}
\end {document}